\documentclass[12pt,a4paper]{article}
\usepackage{ifthen} % for conditional statements
\newboolean{pdflatex}
\setboolean{pdflatex}{true} % False for eps figures 

\newboolean{articletitles}
\setboolean{articletitles}{true} % False removes titles in references

\newboolean{uprightparticles}
\setboolean{uprightparticles}{false} %True for upright particle symbols

\newboolean{inbibliography}
\setboolean{inbibliography}{false} %True once you enter the bibliography

\usepackage{microtype}
\usepackage{lineno}  % for line numbering during review
\usepackage{xspace} % To avoid problems with missing or double spaces after
\usepackage{caption} %these three command get the figure and table captions automatically small

\usepackage{graphicx}  % to include figures (can also use other packages)
\usepackage{color}
\usepackage{colortbl}
\graphicspath{{./figs/}} % Make Latex search fig subdir for figures

\usepackage{amsmath} % Adds a large collection of math symbols
\usepackage{amssymb}
\usepackage{amsfonts}
\usepackage{upgreek} % Adds in support for greek letters in roman typeset

\newcommand*\patchAmsMathEnvironmentForLineno[1]{%
\expandafter\let\csname old#1\expandafter\endcsname\csname #1\endcsname
\expandafter\let\csname oldend#1\expandafter\endcsname\csname
end#1\endcsname
 \renewenvironment{#1}%
   {\linenomath\csname old#1\endcsname}%
   {\csname oldend#1\endcsname\endlinenomath}%
}
\newcommand*\patchBothAmsMathEnvironmentsForLineno[1]{%
  \patchAmsMathEnvironmentForLineno{#1}%
  \patchAmsMathEnvironmentForLineno{#1*}%
}
\AtBeginDocument{%
\patchBothAmsMathEnvironmentsForLineno{equation}%
\patchBothAmsMathEnvironmentsForLineno{align}%
\patchBothAmsMathEnvironmentsForLineno{flalign}%
\patchBothAmsMathEnvironmentsForLineno{alignat}%
\patchBothAmsMathEnvironmentsForLineno{gather}%
\patchBothAmsMathEnvironmentsForLineno{multline}%
\patchBothAmsMathEnvironmentsForLineno{eqnarray}%
}

\usepackage{hyperref}    % Hyperlinks in references
\usepackage[all]{hypcap} % Internal hyperlinks to floats.

\def\lhcb {\mbox{LHCb}\xspace}

\def\cleo   {\mbox{CLEO}\xspace}

\def\MagUp {\mbox{\em Mag\kern -0.05em Up}\xspace}

\ifthenelse{\boolean{uprightparticles}}%
{
 
 \def\Pgamma      {\ensuremath{\upgamma}\xspace}

 \def\Peta        {\ensuremath{\upeta}\xspace}

 \def\Ppi         {\ensuremath{\uppi}\xspace}

 \def\Ppsi        {\ensuremath{\uppsi}\xspace}

 \def\PDelta      {\ensuremath{\Delta}\xspace}                 
 \def\PXi      {\ensuremath{\Xi}\xspace}                 
 \def\PLambda      {\ensuremath{\Lambda}\xspace}                 
 \def\PSigma      {\ensuremath{\Sigma}\xspace}                 
 \def\POmega      {\ensuremath{\Omega}\xspace}                 
 \def\PUpsilon      {\ensuremath{\Upsilon}\xspace}                 
 
 \def\PB      {\ensuremath{\mathrm{B}}\xspace}                 
                  
 \def\PD      {\ensuremath{\mathrm{D}}\xspace}

 \def\PJ      {\ensuremath{\mathrm{J}}\xspace}                 
 \def\PK      {\ensuremath{\mathrm{K}}\xspace}

 \def\PW      {\ensuremath{\mathrm{W}}\xspace}

 \def\Pb      {\ensuremath{\mathrm{b}}\xspace}                 
 \def\Pc      {\ensuremath{\mathrm{c}}\xspace}                 
                  
 \def\Pe      {\ensuremath{\mathrm{e}}\xspace}

 \def\Pi      {\ensuremath{\mathrm{i}}\xspace}

 \def\Ps      {\ensuremath{\mathrm{s}}\xspace}

}
{
 
 \def\Pgamma      {\ensuremath{\gamma}\xspace}

 \def\Peta        {\ensuremath{\eta}\xspace}

 \def\Ppi         {\ensuremath{\pi}\xspace}

 \def\Ppsi        {\ensuremath{\psi}\xspace}                 
                  
 \mathchardef\PDelta="7101
 \mathchardef\PXi="7104
 \mathchardef\PLambda="7103
 \mathchardef\PSigma="7106
 \mathchardef\POmega="710A
 \mathchardef\PUpsilon="7107
                  
 \def\PB      {\ensuremath{B}\xspace}                 
                  
 \def\PD      {\ensuremath{D}\xspace}

 \def\PJ      {\ensuremath{J}\xspace}                 
 \def\PK      {\ensuremath{K}\xspace}

 \def\PW      {\ensuremath{W}\xspace}

 \def\Pb      {\ensuremath{b}\xspace}                 
 \def\Pc      {\ensuremath{c}\xspace}                 
                  
 \def\Pe      {\ensuremath{e}\xspace}

 \def\Pi      {\ensuremath{i}\xspace}

 \def\Ps      {\ensuremath{s}\xspace}

}

\makeatletter
\ifcase \@ptsize \relax% 10pt
  \newcommand{\miniscule}{\@setfontsize\miniscule{4}{5}}% \tiny: 5/6
\or% 11pt
  \newcommand{\miniscule}{\@setfontsize\miniscule{5}{6}}% \tiny: 6/7
\or% 12pt
  \newcommand{\miniscule}{\@setfontsize\miniscule{5}{6}}% \tiny: 6/7
\fi
\makeatother

\DeclareRobustCommand{\optbar}[1]{\shortstack{{\miniscule (\rule[.5ex]{1.25em}{.18mm})}
  \\ [-.7ex] $#1$}}

\def\epem       {{\ensuremath{\Pe^+\Pe^-}}\xspace}
\def\g      {{\ensuremath{\Pgamma}}\xspace}

\def\Wpm    {{\ensuremath{\PW^\pm}}\xspace}

\def\squark    {{\ensuremath{\Ps}}\xspace}

\def\cquark    {{\ensuremath{\Pc}}\xspace}

\def\bquark    {{\ensuremath{\Pb}}\xspace}

\def\pion   {{\ensuremath{\Ppi}}\xspace}

\def\pip    {{\ensuremath{\pion^+}}\xspace}
\def\pim    {{\ensuremath{\pion^-}}\xspace}

\def\kaon    {{\ensuremath{\PK}}\xspace}
  \def\Kbar    {{\kern 0.2em\overline{\kern -0.2em \PK}{}}\xspace}

\def\KorKbar    {\kern 0.18em\optbar{\kern -0.18em K}{}\xspace}
\def\Kz      {{\ensuremath{\kaon^0}}\xspace}

\def\Kp      {{\ensuremath{\kaon^+}}\xspace}
\def\Km      {{\ensuremath{\kaon^-}}\xspace}
\def\Kpm     {{\ensuremath{\kaon^\pm}}\xspace}

\newcommand{\etapr}{\ensuremath{\Peta^{\prime}}\xspace}

  \def\Dbar    {{\kern 0.2em\overline{\kern -0.2em \PD}{}}\xspace}

\def\DorDbar    {\kern 0.18em\optbar{\kern -0.18em D}{}\xspace}

\def\B       {{\ensuremath{\PB}}\xspace}
\def\Bbar    {{\ensuremath{\kern 0.18em\overline{\kern -0.18em \PB}{}}}\xspace}

\def\BorBbar    {\kern 0.18em\optbar{\kern -0.18em B}{}\xspace}
\def\Bz      {{\ensuremath{\B^0}}\xspace}

\def\Bpm     {{\ensuremath{\B^\pm}}\xspace}

\def\Bs      {{\ensuremath{\B^0_\squark}}\xspace}
\def\Bsb     {{\ensuremath{\Bbar{}^0_\squark}}\xspace}

\def\jpsi     {{\ensuremath{{\PJ\mskip -3mu/\mskip -2mu\Ppsi\mskip 2mu}}}\xspace}

  \def\Y#1S{\ensuremath{\PUpsilon{(#1S)}}\xspace}% no space before {...}!

\def\FourS {{\Y4S}}

\def\Lbar        {{\ensuremath{\kern 0.1em\overline{\kern -0.1em\PLambda}}}\xspace}
\def\LorLbar    {\kern 0.18em\optbar{\kern -0.18em \PLambda}{}\xspace}

\def\BF         {{\ensuremath{\cal B}}\xspace}

\newcommand{\decay}[2]{\ensuremath{#1\!\to #2}\xspace}         % {\Pa}{\Pb \Pc}

\def\to                 {\ensuremath{\rightarrow}\xspace}

\def\CP                {{\ensuremath{C\!P}}\xspace}

\newcommand{\ACP}{{\ensuremath{{\cal A}^{\CP}}}\xspace}

\def\AT#1     {\ensuremath{A_{\mathrm{T}}^{#1}}\xspace}           % 2

\def\C#1      {\ensuremath{\mathcal{C}_{#1}}\xspace}                       % 9
\def\Cp#1     {\ensuremath{\mathcal{C}_{#1}^{'}}\xspace}                    % 7
\def\Ceff#1   {\ensuremath{\mathcal{C}_{#1}^{\mathrm{(eff)}}}\xspace}        % 9  
\def\Cpeff#1  {\ensuremath{\mathcal{C}_{#1}^{'\mathrm{(eff)}}}\xspace}       % 7
\def\Ope#1    {\ensuremath{\mathcal{O}_{#1}}\xspace}                       % 2
\def\Opep#1   {\ensuremath{\mathcal{O}_{#1}^{'}}\xspace}                    % 7

\newcommand{\tev}{\ifthenelse{\boolean{inbibliography}}{\ensuremath{~T\kern -0.05em eV}\xspace}{\ensuremath{\mathrm{\,Te\kern -0.1em V}}}\xspace}
\newcommand{\gev}{\ensuremath{\mathrm{\,Ge\kern -0.1em V}}\xspace}
\newcommand{\mev}{\ensuremath{\mathrm{\,Me\kern -0.1em V}}\xspace}
\newcommand{\kev}{\ensuremath{\mathrm{\,ke\kern -0.1em V}}\xspace}
\newcommand{\ev}{\ensuremath{\mathrm{\,e\kern -0.1em V}}\xspace}
\newcommand{\gevc}{\ensuremath{{\mathrm{\,Ge\kern -0.1em V\!/}c}}\xspace}
\newcommand{\mevc}{\ensuremath{{\mathrm{\,Me\kern -0.1em V\!/}c}}\xspace}
\newcommand{\gevcc}{\ensuremath{{\mathrm{\,Ge\kern -0.1em V\!/}c^2}}\xspace}
\newcommand{\gevgevcccc}{\ensuremath{{\mathrm{\,Ge\kern -0.1em V^2\!/}c^4}}\xspace}
\newcommand{\mevcc}{\ensuremath{{\mathrm{\,Me\kern -0.1em V\!/}c^2}}\xspace}

\def\mm   {\ensuremath{\rm \,mm}\xspace}

\def\invfb   {\ensuremath{\mbox{\,fb}^{-1}}\xspace}

\def\gsim{{~\raise.15em\hbox{$>$}\kern-.85em
          \lower.35em\hbox{$\sim$}~}\xspace}
\def\lsim{{~\raise.15em\hbox{$<$}\kern-.85em
          \lower.35em\hbox{$\sim$}~}\xspace}

\def\pt         {\mbox{$p_{\rm T}$}\xspace}

\def\tell1  {TELL1\xspace}
\def\ukl1   {UKL1\xspace}

\usepackage{cite} % Allows for ranges in citations
\usepackage{mciteplus}

\usepackage{longtable} % only for template; not usually to be used in PAPERs

\begin{document}
%%%%%%%%%%%%%%%%%%%%%%%%%
%%%%% Title     %%%%%%%%%
%%%%%%%%%%%%%%%%%%%%%%%%%
\renewcommand{\thefootnote}{\fnsymbol{footnote}}
\setcounter{footnote}{1}

% %%%%%%% CHOOSE TITLE PAGE--------
%\onecolumn
% \input{title-LHCb-ANA}
%\input{title-LHCb-CONF}
% $Id: title-LHCb-PAPER.tex 61931 2014-10-14 09:51:37Z roldeman $
% ===============================================================================
% Purpose: LHCb-PAPER journal paper title page template
% Author: 
% Created on: 2010-09-25
% ===============================================================================

%%%%%%%%%%%%%%%%%%%%%%%%%
%%%%%  TITLE PAGE  %%%%%%
%%%%%%%%%%%%%%%%%%%%%%%%%
\begin{titlepage}
\pagenumbering{roman}

% Header ---------------------------------------------------
\vspace*{-1.5cm}
\centerline{\large EUROPEAN ORGANIZATION FOR NUCLEAR RESEARCH (CERN)}
\vspace*{1.5cm}
\hspace*{-0.5cm}
\begin{tabular*}{\linewidth}{lc@{\extracolsep{\fill}}r}
%\ifthenelse{\boolean{pdflatex}}% Logo format choice
%{\vspace*{-2.7cm}\mbox{\!\!\!\includegraphics[width=.14\textwidth]{lhcb-logo.pdf}} & &}%
%{\vspace*{-1.2cm}\mbox{\!\!\!\includegraphics[width=.12\textwidth]{lhcb-logo.eps}} & &}%
\\
 & & CERN-PH-EP-2015-073 \\  % ID 
 & & LHCb-PAPER-2014-065 \\  % ID 
 & & 25 March 2015 \\ % Date - Can also hardwire e.g.: 23 March 2010
 & & \\
% not in paper \hline
\end{tabular*}

\vspace*{2.0cm}

% Title --------------------------------------------------
{\bf\boldmath\huge
\begin{center}
Observation of the $B^0_s\to\etapr\etapr$ decay
%and measurement of \decay{\Bpm}{\etapr\Kpm}  and \decay{\Bpm}{\phi\Kpm} charge asymmetries
\end{center}
}

\vspace*{2.0cm}

% Authors -------------------------------------------------
\begin{center}
%In the footnote, replace 'paper' by 'letter' in case of submission to PRL or PLB 
The LHCb collaboration\footnote{Authors are listed at the end of this Letter.}
\end{center}

\vspace{\fill}

% Abstract -----------------------------------------------

\begin{abstract}
\noindent
The first observation of the $B^0_s\to\eta'\eta'$ decay is reported. The study is based on a sample
of proton-proton collisions corresponding to $3.0$\invfb of integrated luminosity collected with the LHCb detector.
The significance of the signal is $6.4$ standard deviations. The branching fraction is measured to be
$[3.31 \pm 0.64\,{\rm (stat)} \pm 0.28\,{\rm (syst)} \pm 0.12\,{\rm (norm)}]\times10^{-5}$, where the third uncertainty
comes from the \decay{\Bpm}{\etapr\Kpm}  branching fraction that is used as a normalisation.
In addition, the charge asymmetries of \decay{\Bpm}{\etapr\Kpm}  and \decay{\Bpm}{\phi\Kpm},
which are control channels, are measured to be $(-0.2 \pm1.3)\%$ and $(+1.7\pm1.3)\%$, respectively.
All results are consistent with theoretical expectations.
\end{abstract}

\vspace*{2.0cm}

\begin{center}
Submitted to Phys.~Rev.~Lett. 
\end{center}

\vspace{\fill}

{\footnotesize 
\centerline{\copyright~CERN on behalf of the \lhcb collaboration, licence \href{http://creativecommons.org/licenses/by/4.0/}{CC-BY-4.0}.}}
\vspace*{2mm}

\end{titlepage}

%%%%%%%%%%%%%%%%%%%%%%%%%%%%%%%%
%%%%%  EOD OF TITLE PAGE  %%%%%%
%%%%%%%%%%%%%%%%%%%%%%%%%%%%%%%%

%  empty page follows the title page ----
\newpage
\setcounter{page}{2}
\mbox{~}

\cleardoublepage

%\twocolumn
% %%%%%%%%%%%%% ---------

\renewcommand{\thefootnote}{\arabic{footnote}}
\setcounter{footnote}{0}

%%%%%%%%%%%%%%%%%%%%%%%%%%%%%%%%
%%%%%  Table of Content   %%%%%%
%%%%%%%%%%%%%%%%%%%%%%%%%%%%%%%%
%%%% Uncomment next 2 lines if desired
%\tableofcontents
%\cleardoublepage

%%%%%%%%%%%%%%%%%%%%%%%%%
%%%%% Main text %%%%%%%%%
%%%%%%%%%%%%%%%%%%%%%%%%%

\pagestyle{plain} % restore page numbers for the main text
\setcounter{page}{1}
\pagenumbering{arabic}

%% Uncomment during review phase. 
%% Comment before a final submission.
%\linenumbers

\newcommand{\ACPraw}{{\ensuremath{{\cal A}^{\CP}_{\rm raw}}}\xspace}
\newcommand{\ACPrawk}{{\ensuremath{{\cal A}^{\CP}_{{\rm raw},k}}}\xspace}
\newcommand{\ACPphys}{{\ensuremath{{\cal A}^{\CP}}}\xspace}

Hadronic  decays of beauty hadrons into final states without charm quarks (charmless decays) 
are suppressed in the Standard Model of elementary particles. 
They proceed predominantly through $b\to u$ transitions, mediated by the emission of a virtual 
\Wpm boson, 
and
$b\to s$ transitions, mediated by the exchange of a virtual \Wpm boson and a virtual quark. 
The respective ``tree'' and ``penguin'' amplitudes 
are of similar size,
allowing for possible large quantum interference effects 
measurable as charge-parity (\CP)
violating asymmetries.
New particles not described in the Standard Model may contribute with additional 
amplitudes, and therefore affect both the decay rates and the \CP asymmetries~\cite{Zhang:2000ic}.
The $\Bpm\to\etapr\Kpm$ and $\Bz\to\etapr\Kz$ decays,
\footnote{Charge conjugation of
neutral $B^0_{(\squark)}$ mesons is implied throughout this Letter.
The notations \etapr and $\phi$ refer to the $\etapr(958)$
and $\phi(1020)$ mesons, respectively.}
first observed by the \cleo
collaboration~\cite{Behrens:1998dn}, have some of the largest branching fractions among all
charmless hadronic $B$-meson decays~\cite{PDG2014}. Studies of such decays, 
conducted so far mostly at \epem~colliders operating at the \FourS~resonance,
provide accurate measurements of integrated~\cite{Aubert:2009yx, Schumann:2006bg} and
time-dependent~\cite{Aubert:2008ad, Chen:2006nk} \CP-violating asymmetries in charmless
hadronic \Bpm and \Bz meson decays, and are useful to look for deviations from Standard Model predictions. 

Charmless hadronic \Bs decays are poorly known, in particular decays to a pair of unflavoured neutral mesons~\cite{Acciarri:1995bx,Abe:1999ze},
but have been extensively studied in the framework
of QCD factorisation~\cite{Cheng:2009mu, Sun:2002rn,Beneke:2003zv}, perturbative QCD~\cite{Ali:2007ff},
soft collinear effective theory~\cite{Williamson:2006hb},
and flavour SU(3) symmetry~\cite{Cheng:2014rfa}.
The decay \decay{\Bs}{\etapr\etapr} is expected to have a relatively large branching fraction, similar to that of its
SU(3) counterpart \decay{\B}{\etapr K};
predictions range between $14\times 10^{-6}$ and $50\times 10^{-6}$,
and have large uncertainties~\cite{Cheng:2009mu, Sun:2002rn,Ali:2007ff,Williamson:2006hb,Cheng:2014rfa,Beneke:2003zv}.
The $\etapr\etapr$ final state is a pure \CP eigenstate. 
Decays to this final state of \Bs and \Bsb mesons flavour tagged at production
may therefore be used to investigate 
time-dependent \CP asymmetries in a complementary way to the measurements in 
\decay{\Bs}{\phi\phi}~\cite{LHCb-PAPER-2014-026}, but without the need for an angular analysis.

In this Letter we present the first observation of the \decay{\Bs}{\etapr\etapr} decay. 
Its branching fraction is measured using the known 
\decay{\Bpm}{\etapr\Kpm} and \decay{\Bpm}{\phi\Kpm} 
decays as calibration channels.
The \CP asymmetries of the calibration channels are also measured, relatively to the \decay{\Bpm}{\jpsi\Kpm} channel.
All these measurements use proton-proton ($pp$)
collisions corresponding to $3.0$\invfb of integrated luminosity, of which $1.0$ $(2.0)$\invfb was collected in
2011 (2012) at a centre-of-mass energy of $7$ $(8)$\tev with the LHCb detector.

The \lhcb detector~\cite{Alves:2008zz}
 is a single-arm forward
spectrometer at the LHC covering the \mbox{pseudorapidity} range $2<\eta <5$,
designed for the study of particles containing~\bquark or \cquark~quarks. 
The detector includes a high-precision tracking system, two ring-imaging Cherenkov detectors
used to distinguish different types of charged hadrons,
a calorimeter system consisting of scintillating-pad and preshower detectors, an electromagnetic
calorimeter and a hadronic calorimeter, and a muon system.
The trigger~\cite{LHCb-DP-2012-004} consists of a
hardware stage, based on information from the calorimeter and muon
systems, followed by a software stage, which applies event
reconstruction using information from all the detector sub-systems.

Signal $\decay{\Bs}{\etapr\etapr}$, $\decay{\Bpm}{\etapr\Kpm}$, and $\decay{\Bpm}{\phi\Kpm}$ candidates are reconstructed through the decays \decay{\etapr}{\pip\pim\gamma} and \decay{\phi}{\Kp\Km}.
Selection requirements are chosen to be as similar as possible for the three channels and are optimised
for $\decay{\Bs}{\etapr\etapr}$, maximising the figure of merit
$\varepsilon/(a/2 + \sqrt{B})$~\cite{Punzi:2003bu},
where $\varepsilon$ is the efficiency for selecting simulated signal events,
$B$ is the number of background events in the signal region estimated from the mass sidebands,
and $a=5$ is the target significance of the possible signal.
The requirements on the $\phi$ meson and on the charged kaon associated with the $\etapr$ or $\phi$ resonance
in the candidate $\Bpm$ decays, which is referred to as the bachelor kaon, are chosen to minimise the relative
statistical uncertainty on the $\decay{\Bpm}{\phi\Kpm}$ signal yield. 

Charged particles are required to be inconsistent with originating from a primary $pp$ interaction vertex (PV)
and to have a transverse momentum (\pt) with respect to the beam line in excess of $0.25$\gevc,
while bachelor kaons must have $\pt >1.2$\gevc. Particle identification algorithms are applied to
distinguish kaons from pions, and photons from electrons~\cite{LHCb-DP-2014-002}.
Photons are required to have $\pt >0.5$\gevc.
The intermediate \etapr ($\phi$) resonances must have $\pt>1.5(0.5)$\gevc and momentum $p>4$\gevc.
The $\pip\pim$ invariant mass in candidate \etapr decays must exceed $0.56$\gevcc.
$B$-meson candidates must have $\pt>4$\gevc.
Topological variables are used to isolate the signal, such as the angle between the
reconstructed $B$ momentum and the vector pointing from the PV to the $B$ decay vertex
(required to be smaller than $10$ mrad),
and the distance of closest approach to the PV of the $B$ trajectory (required to be less than $0.04$\mm).
Reconstructed invariant masses of the \Bs, \Bpm, \etapr, and $\phi$ candidates are required to be in the ranges $5000<m_{\Bs}<5600\mevcc$, $5000<m_{\Bpm}<5500\mevcc$,  $880<m_{\etapr}<1040\mevcc$ and $1000<m_{\phi}<1050\mevcc$, respectively.
Only candidates with a well reconstructed $B$ decay vertex are retained; in the events with multiple candidates
($\lesssim 5$\%), the candidate with the smallest vertex $\chi^2$ is kept.

The $\decay{\Bs}{\etapr\etapr}$ signal yield is determined from a multidimensional unbinned
extended maximum likelihood fit to the sample of $\decay{\Bs}{\etapr\etapr}$
and $\decay{\Bpm}{\etapr\Kpm}$ candidates, using the combined $\sqrt{s}=7$ and~$8$\tev data sets.
The likelihood is written as
${\cal L} = {\textstyle \exp(-\sum_j N_j)} \, \prod_i^N
{\left( \textstyle \sum_j N_j P_j^i\right)}$,
where $N_j$ is the yield of fit component~$j$ (signal or backgrounds), $P_j^i$ is the probability
of event $i$ for component~$j$, and $N$ is the total number of events.
The probabilities $P_j^i$ are expressed as products of
probability density functions (PDF) for the invariant masses used as observables in the fit:
the $\etapr\etapr$ invariant mass ($m_{\etapr\etapr}$), the two randomly ordered $\pip\pim\g$
invariant masses ($m_{(\pi\pi\g)_1}$ and $m_{(\pi\pi\g)_2}$) of the \decay{\Bs}{\etapr\etapr} candidates,
and the $\etapr\Kpm$ and $\pip\pim\g$ invariant masses ($m_{\etapr K}$ and $m_{\pi\pi\g}$) of the \decay{\Bpm}{\etapr\Kpm} candidates.
In the reconstruction of the $\pip\pim\g$ candidates, the known \etapr mass~\cite{PDG2014} is applied as a constraint to calculate the $m_{\etapr\etapr}$ and $m_{\etapr K}$ variables.

The \decay{\Bpm}{\etapr \Kpm} sample is described with three components: the signal, and two
combinatorial background components with and without an $\etapr$ resonance in the decay chain.
The \decay{\Bs}{\etapr \etapr} sample is modelled with seven components, three of which are significant:
the signal, the combinatorial background and partially reconstructed $b$-hadron decays without
$\etapr$ resonances in the final state.
The remaining backgrounds, for which the event yields are found to be consistent with zero, consist of two combinatorial and two partially reconstructed components, each involving only one resonant $\pip\pim\g$ candidate.
 
All the PDFs that peak at the $B$ or $\etapr$ mass
are modelled by Crystal Ball (CB) functions~\cite{Skwarnicki:1986xj}
modified such that both the high- and low-mass tails follow power laws and account for non Gaussian reconstruction effects. 
The parameters used for the description of $m_{\etapr K}$ for the $\Bpm$ signal are free in
the fit, while all the parameters of $m_{\etapr\etapr}$ for the $\Bs$ signal PDF are determined from simulation,
except the CB width. The ratio of the CB widths in $m_{\etapr K}$ and $m_{\etapr\etapr}$ is fixed
to that measured in simulation.
The partially reconstructed background is described with an ARGUS function~\cite{Argus} convolved with
a Gaussian resolution function of the same width as the corresponding signal PDF,
while the combinatorial background is modelled with a linear function.
A common CB function is used for modelling all the $\etapr$ resonances, with mean and width free in the fit,
while tail parameters are determined from simulation. The mass distribution
of \etapr candidates from random combinations is modelled with an empirical quadratic function.

\begin{figure}[ht]
\centering
\includegraphics[width=0.7\linewidth]{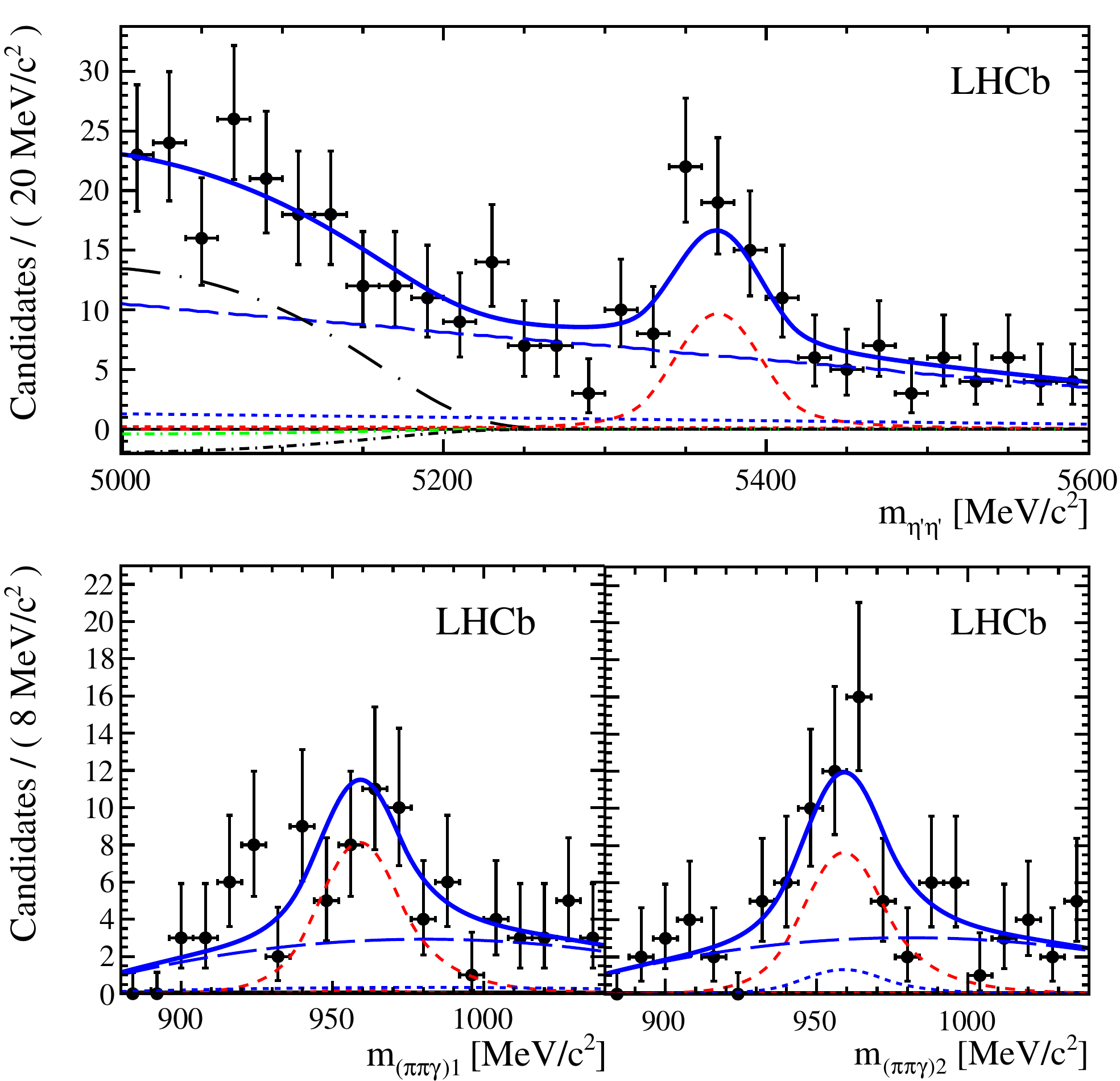}
\caption{Distributions of the (top) $\etapr\etapr$, (bottom) $(\pi\pi\g)_1$ and $(\pi\pi\g)_2$
invariant masses for the $\decay{\Bs}{\etapr\etapr}$ candidates with fit results overlaid.
The $(\pi\pi\g)_1$ and $(\pi\pi\g)_2$ mass distributions are shown for the candidates with
an $\etapr\etapr$ invariant mass within three standard deviations of the \Bs mass. 
The components are the following: (dashed red curves) $\decay{\Bs}{\etapr\etapr}$ signal, 
(long-dashed blue curves) combinatorial background without an $\etapr$ resonance in the final state,
(dot-long-dashed black curves) partially reconstructed background without an $\etapr$ resonance,
(short-dashed red, short-dashed blue curves) combinatorial background with one $\etapr$ resonance,
and (dot-dashed green, dot-dashed black curves) partially reconstructed background with one $\etapr$ resonance.
The total fit function is shown as the solid blue curves. 
}
\label{fig:EtapEtap}
\end{figure}

\begin{figure}[ht]
\centering
\begin{center}
\includegraphics[width=0.7\linewidth]{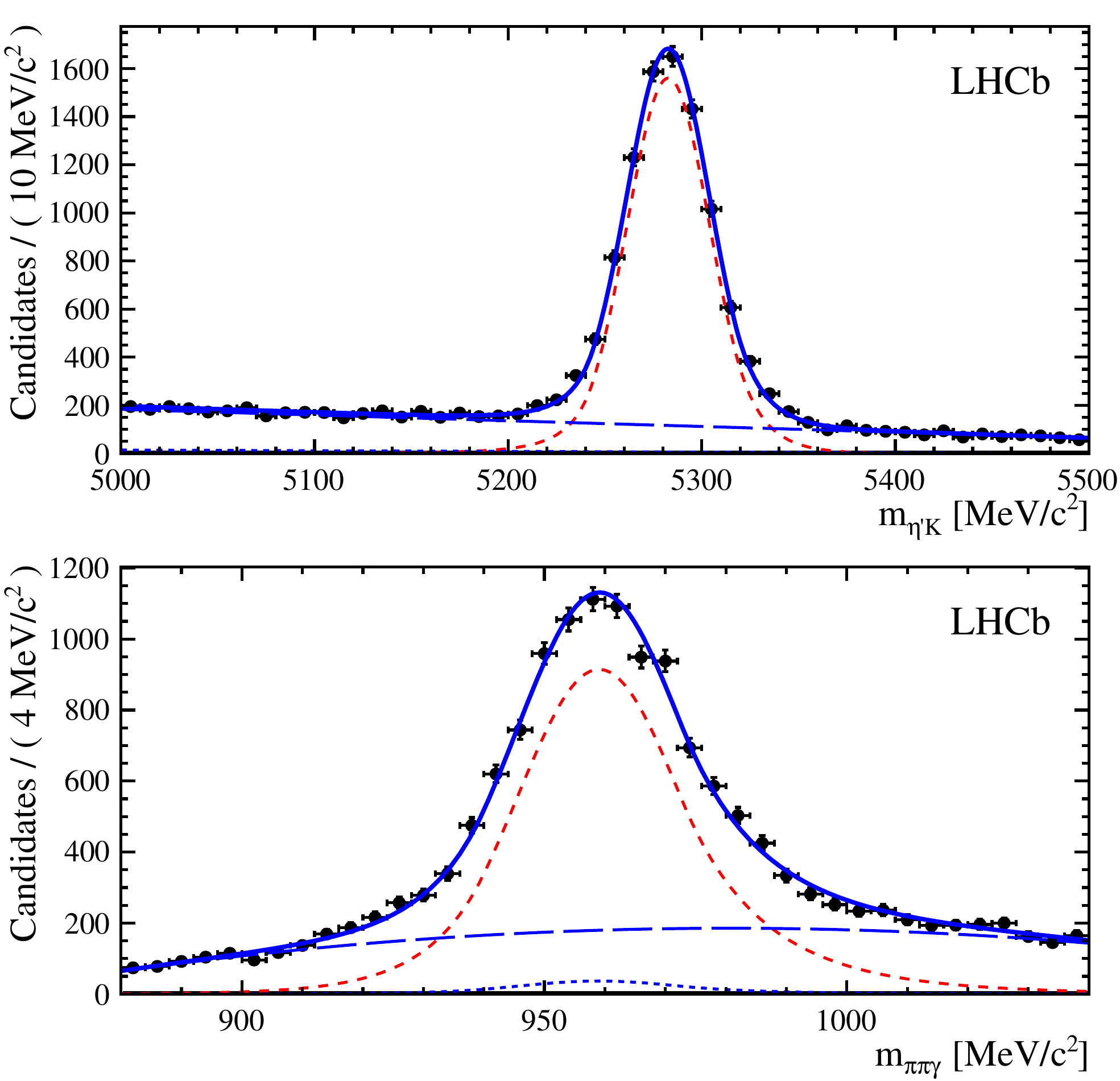}
\end{center}
\caption{Distributions of the (top) $\etapr\Kpm$ and (bottom) $\pip\pim\gamma$
invariant masses for the $\decay{\Bpm}{\etapr\Kpm}$ candidates with fit results overlaid.
The components are the following: (dashed red curves) $\decay{\Bpm}{\etapr\Kpm}$ signal, 
(long-dashed blue curves) combinatorial background without an $\etapr$ resonance in the final state,
and (dotted blue curves) combinatorial background with an $\etapr$ resonance.
The total fit function is shown as the solid blue curves.}
\label{fig:EtapK}
\end{figure}

We observe $36.4 \pm 7.8\,({\rm stat}) \pm 1.6\,({\rm syst})$ \decay{\Bs}{\etapr \etapr} decays corresponding to a significance of $6.4$ standard deviations, including both statistical and systematic uncertainties, discussed later.
The significance is computed using Wilks' theorem~\cite{wilks1938}, and is scaled by the ratio of the statistical over
the total uncertainties.
The measured $\decay{\Bpm}{\etapr\Kpm}$ yield is $8672 \pm 114$, where the uncertainty is statistical only.
Mass distributions are shown in Figs.~\ref{fig:EtapEtap} and~\ref{fig:EtapK}, with the fit results overlaid.

To measure the ratio of the branching fractions
$\BF(\decay{\Bs}{\etapr\etapr})/\BF(\decay{\Bpm}{\etapr\Kpm})$, the fit is repeated taking into account the different
reconstruction efficiencies in the $7$ and $8$\tev data sets.
The four background components with yields consistent with zero are neglected in this case. The common parameters
between the $7$ and $8$\tev data sets are the shape parameters and the ratio of branching fractions.
The ratio of branching fractions is related to the ratio of yields according to
\begin{align}
\label{eq:FormulaBF}
& \frac{\BF(\decay{\Bs}{\etapr\etapr})}{ \BF(\decay{\Bpm}{\etapr\Kpm}) }=
 \frac{f_d}{f_s}\times \frac{1}{\BF(\decay{\etapr}{\pip\pim\g})} \nonumber \\ &
 \times\frac{ N_l(\decay{\Bs}{\etapr\etapr})}{ N_l(\decay{\Bpm}{\etapr\Kpm})}  \times\frac{\varepsilon_l(\decay{\Bpm}{\etapr\Kpm})}{\varepsilon_l(\decay{\Bs}{\etapr\etapr})} \,,
\end{align}
where the subscript $l$ indicates the $7$\tev or $8$\tev data set, the ratio of probabilities for a $b$ quark to produce
a \Bs or \Bpm meson is $f_s/f_d = 0.259 \pm 0.015$~\cite{fsfd}, and the branching fraction of the \etapr decay is
$\BF(\decay{\etapr}{\pip\pim\g})=0.291 \pm 0.005$~\cite{PDG2014}. The ratio of efficiencies for
reconstructing the normalisation and signal decay channels $\varepsilon_l(\decay{\Bpm}{\etapr\Kpm})/\varepsilon_l(\decay{\Bs}{\etapr\etapr})$ is determined from control samples (particle identification, photon reconstruction and hardware trigger on the signal) and simulation
to be $8.46 \pm 0.35$ for the $7$\tev and $7.85 \pm 0.26$ for the $8$\tev data sets, including all experimental systematic uncertainties.
The largest uncertainty in the determination of the efficiency ratio comes from
the photon reconstruction efficiency. 
 This efficiency is measured using $B^\pm \to J/\psi K^{*\pm}$ decays followed by $J/\psi \to \mu^+\mu^-$ and $K^{*\pm} \to K^\pm\pi^0 \to  K^\pm\gamma\gamma$\cite{LHCb:2012cw},
and a crosscheck is provided by the measurement of the ratio of branching
fractions of the \decay{\Bpm}{\etapr\Kpm} and \decay{\Bpm}{\phi\Kpm} control channels.
The result of Eq.~(\ref{eq:FormulaBF}) is then
\begin{equation}
\frac{\BF(\decay{\Bs}{\etapr\etapr})}{ \BF(\decay{\Bpm}{\etapr\Kpm}) }= 0.47 \pm 0.09\,({\rm stat}) \pm 0.04\,({\rm syst}) \nonumber \,.
\end{equation}

Contributions to the systematic uncertainties 
are summed in quadrature leading to a total systematic uncertainty on the \decay{\Bs}{\etapr \etapr} signal yield (ratio of branching fractions) of $1.6$ ($0.041$).
The uncertainties on $f_s/f_d$, \BF(\decay{\etapr}{\pip\pim\g}),
$\varepsilon_l(\decay{\Bpm}{\etapr\Kpm})/\varepsilon_l(\decay{\Bs}{\etapr\etapr})$,
and on the values of fit model parameters fixed from simulation, lead to a systematic
uncertainty of $0.7$ ($0.038$), while a variation of the PDF models leads to an uncertainty of $1.4$ ($0.007$).
The fit bias, evaluated in simulation, is consistent with zero, and its statistical uncertainty of $0.4$ ($0.005$) is applied
as a systematic uncertainty.
Finally, an uncertainty of $0.014$ is assigned to account for the neglected background components
in the branching-fraction fit.

Using the known value $\BF(\decay{\Bpm}{\etapr\Kpm}) = (7.06\pm 0.25)\times 10^{-5}$~\cite{PDG2014},
the branching fraction is measured to be
$\mathcal{B}(B^0_s\to\eta'\eta') = [3.31\, \pm 0.64\,({\rm stat}) \pm 0.28\,({\rm syst}) \pm 0.12\,({\rm norm})]\times 10^{-5}$,
where the third uncertainty comes from the \decay{\Bpm}{\etapr\Kpm}  branching fraction.

The $\decay{\Bpm}{\etapr\Kpm}$  and $\decay{\Bpm}{\phi\Kpm}$ charge asymmetries, $\ACPphys \equiv (\Gamma^- -\Gamma^+)/(\Gamma^- +\Gamma^+)$,
where $\Gamma^{\pm}$ is the partial width of the \Bpm meson, are
determined using the strategy adopted in Ref.~\cite{2013lja}.
For these measurements, we consider either events triggered on signal candidates (TOS events)
or events triggered at the hardware stage independently of the signal candidate (non-TOS events).
The raw asymmetries \ACPraw are obtained from a fit to the positively  and negatively charged candidates,
and each subsample is further split into TOS and non-TOS events,
to account for trigger-dependent detection asymmetries~\cite{2013lja}.
For each channel a two-dimensional fit of the mass  distributions of the $B$ candidate
and its neutral daughter is performed, and the four samples are fitted simultaneously.

The model in the $\decay{\Bpm}{\etapr\Kpm}$ fit is the same as that used in the simultaneous $\decay{\Bs}{\etapr\etapr}$ and $\decay{\Bpm}{\etapr\Kpm}$ fit; the $\decay{\Bpm}{\phi\Kpm}$ model mirrors that of Ref.~\cite{2013lja}, except for the $m_{\phi K}$ signal PDF, which is described by the sum of a CB function and a Gaussian function.
\begin{figure}[t]
\centering
\includegraphics[width=0.7\linewidth]{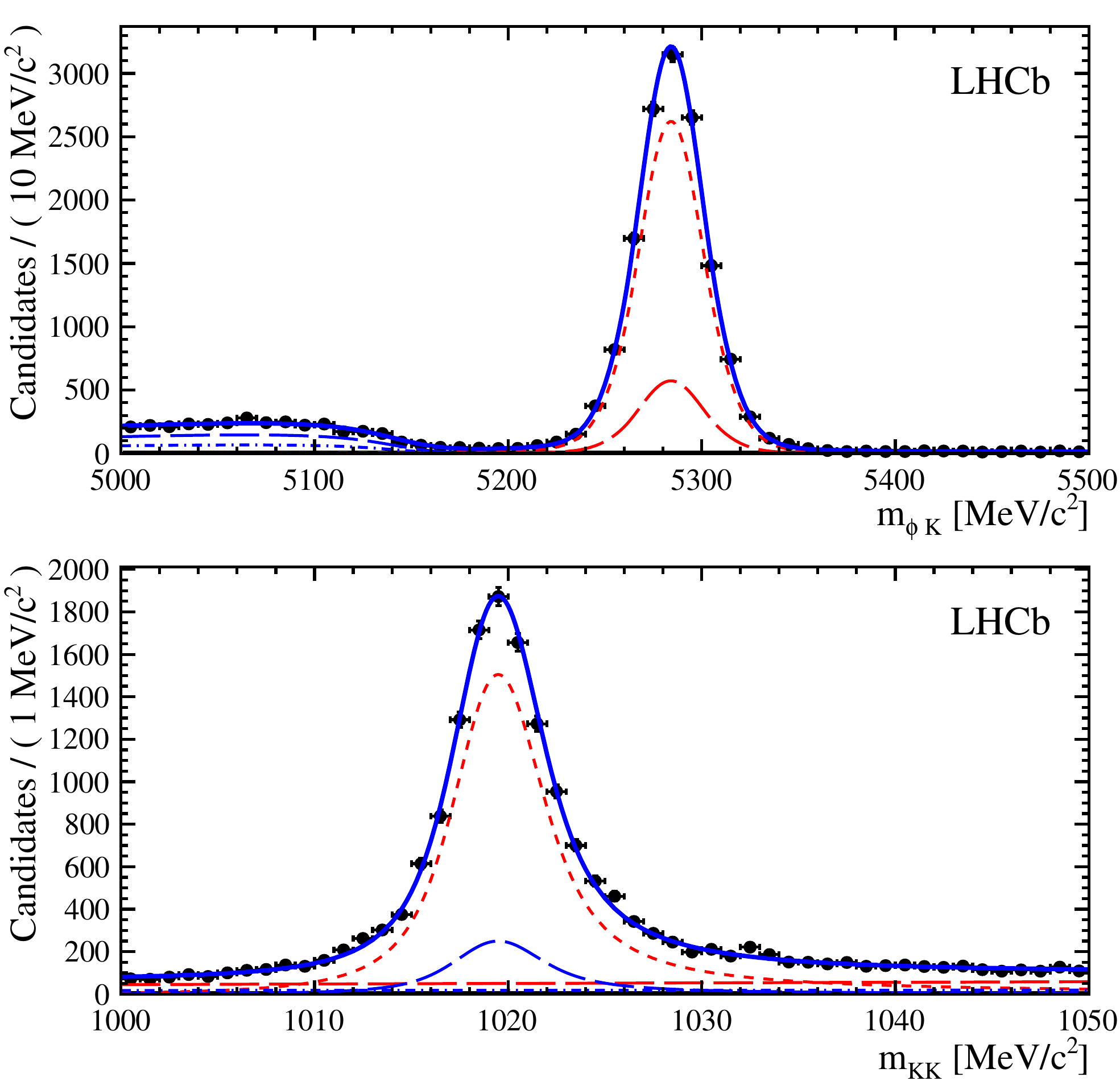}
\caption{Distributions of the (top) $\phi\Kpm$ and (bottom) $\Kp\Km$ invariant masses
for the $\decay{\Bpm}{\phi\Kpm}$ candidates with fit results overlaid
The components are the following:
(dashed red curves) \decay{\Bpm}{\phi\Kpm} signal, 
(long-dashed red curves) non-resonant \decay{\Bpm}{\Kp\Km\Kpm} background,
partially reconstructed $b$-hadron background (long-dashed blue curves) with or
(dot-dashed blue curves) without a $\phi$ resonance in the decay chain.
The combinatorial background with or without a $\phi$ resonance is too small to be visible.
The total fit function is shown as the solid blue curves.}
\label{fig:PhiK}
\end{figure}
The shape parameters used to describe
the peaking and partially reconstructed component PDFs are shared, while independent parameters are
used for the combinatorial component PDFs in the trigger sub-samples.
In both fits, each component has four yields parametrised as
$N^{\pm}_{k} = N_{k} (1\mp \ACPrawk)/2$, 
where $N_k$ and \ACPrawk are the yield and the raw asymmetry in each trigger category~$k$, respectively.

The observed \decay{\Bpm}{\etapr\Kpm} (\decay{\Bpm}{\phi\Kpm}) raw asymmetries and their statistical uncertainties are
$-0.019\pm0.014$ ($+0.003\pm0.014$) and $-0.027\pm0.020$ ($-0.011\pm0.018$) for the TOS and non-TOS
categories, respectively, the fraction of event in the TOS category being $63.8\%$ ($60.1\%$).
The fitted mass spectra for the $\decay{\Bpm}{\phi\Kpm}$ candidates are shown in Fig.~\ref{fig:PhiK}.

In order to determine the \CP asymmetry, the raw asymmetry is corrected for the $\Bpm$ production
asymmetry in $pp$ collisions, ${\cal A}_{\rm P}$, and for the bachelor $\Kpm$ detection asymmetry
due to interactions with the detector matter, ${\cal A}_{{\rm D},k}$. Under the assumption that these asymmetries
are small, \ACPrawk is related to the \CP asymmetry \ACPphys as
$\ACPrawk = \ACPphys +{\cal A}_{{\rm D},k} + {\cal A}_{\rm P}$.
Because the \CP-violating asymmetry in \decay{\Bpm}{\jpsi\Kpm} is known precisely~\cite{PDG2014},
the raw asymmetry in this decay is used to determine the sum of the detection and production asymmetries.
The raw asymmetry of the $\Bpm\to\jpsi\Kpm$  decay is measured in a simultaneous fit to the $m_{\jpsi\Kpm}$ distributions of the positively  and negatively charged candidates selected with similar criteria as for the signal modes. Independent fits are performed for events belonging to each trigger category. The fit model consists of a peaking signal component, described with the sum of a CB function and a Gaussian function, and a linear combinatorial background component.

The raw $\Bpm\to\jpsi\Kpm$ asymmetries, $-0.020\pm0.004\,{\rm (stat)} $ and $-0.011\pm0.003\,{\rm (stat)} $ for the TOS and non-TOS categories, respectively, are subtracted from the \decay{\Bpm}{\etapr\Kpm} or \decay{\Bpm}{\phi\Kpm} raw asymmetries for each trigger category. The weighted average of these asymmetry differences, $\Delta \ACP=\ACP-\ACP(\decay{\Bpm}{\jpsi\Kpm})$, is computed with weights given by the fractions of signal events in each of the two categories.  The resulting asymmetry differences are $\Delta \ACP(\decay{\Bpm}{\etapr\Kpm}) = -0.005\pm0.012\,{\rm (stat)} $ and $\Delta \ACP(\decay{\Bpm}{\phi\Kpm}) = +0.014\pm0.011\,{\rm (stat)}$.

Three significant sources of systematic uncertainties are identified. The first accounts for the effect
of the mass shape modelling, leading to an uncertainty on $\Delta\ACP$ of $0.021\times10^{-2}$
($0.20\times10^{-2}$) for the \decay{\Bpm}{\etapr\Kpm} (\decay{\Bpm}{\phi\Kpm}) channel.
To account for the different kinematic properties between the signal channels and the \decay{\Bpm}{\jpsi\Kpm} channel,
$\Delta\ACP$ is measured in three independent subsamples selected according to the transverse
momentum of the bachelor kaon, and their average, weighted by the number of events in each subsample, is computed.
The difference from the result obtained in the default fit is $0.018\times 10^{-2}$ ($0.08\times 10^{-2}$), which is assigned as a systematic uncertainty.
Finally, the \CP measurements are repeated applying a geometrical requirement~\cite{2013lja} to suppress possible detector edge effects. The associated systematic uncertainty is $0.13\times 10^{-2}$ ($0.05\times 10^{-2}$).

Using the known \decay{\Bpm}{\jpsi\Kpm} \CP asymmetry, $\ACP(\decay{\Bpm}{\jpsi\Kpm}) = (+0.3 \pm 0.6)\times 10^{-2}$~\cite{PDG2014},  the asymmetries are measured to be
$\ACP(\decay{\Bpm}{\etapr\Kpm}) = [-0.2\pm 1.2\,({\rm stat}) \pm 0.1\,({\rm syst})\pm 0.6\,({\rm norm})] \times 10^{-2}$
and
$\ACP(\decay{\Bpm}{\phi\Kpm})     = [+1.7\pm 1.1\,({\rm stat}) \pm 0.2\,({\rm syst})\pm 0.6\,({\rm norm})] \times 10^{-2}$,
where the third uncertainty comes from the \decay{\Bpm}{\jpsi\Kpm} \CP asymmetry. These results are compatible with
the hypothesis of \CP symmetry and with the Standard Model predictions~\cite{Li:2006jv,Beneke:2003zv}.

In conclusion, this Letter presents the first observation of the decay $\decay{\Bs}{\etapr\etapr}$,
with a significance of $6.4$~standard deviations, and the most precise measurements of \CP-violating
charge asymmetries in $\decay{\Bpm}{\etapr\Kpm}$ and
$\decay{\Bpm}{\phi\Kpm}$ decays. The latter result supersedes the previous \lhcb measurement~\cite{2013lja}.
The measured $\decay{\Bs}{\etapr\etapr}$ branching fraction,
$\mathcal{B}(B^0_s\to\eta'\eta') = [3.31\, \pm 0.64\,({\rm stat}) \pm 0.28\,({\rm syst})\pm 0.12\,({\rm norm})]\times 10^{-5}$,
agrees with the theoretical predictions. This newly observed \Bs decay channel to a charmless \CP eigenstate 
opens possibilities for further constraining
the Standard Model with time-dependent \CP asymmetry measurements.

\section*{Acknowledgements}
 
\noindent We express our gratitude to our colleagues in the CERN
accelerator departments for the excellent performance of the LHC. We
thank the technical and administrative staff at the LHCb
institutes. We acknowledge support from CERN and from the national
agencies: CAPES, CNPq, FAPERJ and FINEP (Brazil); NSFC (China);
CNRS/IN2P3 (France); BMBF, DFG, HGF and MPG (Germany); INFN (Italy); 
FOM and NWO (The Netherlands); MNiSW and NCN (Poland); MEN/IFA (Romania); 
MinES and FANO (Russia); MinECo (Spain); SNSF and SER (Switzerland); 
NASU (Ukraine); STFC (United Kingdom); NSF (USA).
The Tier1 computing centres are supported by IN2P3 (France), KIT and BMBF 
(Germany), INFN (Italy), NWO and SURF (The Netherlands), PIC (Spain), GridPP 
(United Kingdom).
We are indebted to the communities behind the multiple open 
source software packages on which we depend. We are also thankful for the 
computing resources and the access to software R\&D tools provided by Yandex LLC (Russia).
Individual groups or members have received support from 
EPLANET, Marie Sk\l{}odowska-Curie Actions and ERC (European Union), 
Conseil g\'{e}n\'{e}ral de Haute-Savoie, Labex ENIGMASS and OCEVU, 
R\'{e}gion Auvergne (France), RFBR (Russia), XuntaGal and GENCAT (Spain), Royal Society and Royal
Commission for the Exhibition of 1851 (United Kingdom).

\newpage

\addcontentsline{toc}{section}{References}
%\setboolean{inbibliography}{true}
%\bibliographystyle{LHCb}
%\bibliography{main,LHCb-PAPER,LHCb-CONF,LHCb-DP,LHCb-TDR}

\begin{mcitethebibliography}{10}
\mciteSetBstSublistMode{n}
\mciteSetBstMaxWidthForm{subitem}{\alph{mcitesubitemcount})}
\mciteSetBstSublistLabelBeginEnd{\mcitemaxwidthsubitemform\space}
{\relax}{\relax}
\bibitem{Zhang:2000ic}
D.~Zhang, Z.~Xiao, and C.~S. Li,
  \ifthenelse{\boolean{articletitles}}{\emph{{Branching ratios and CP-violating
  asymmetries of $B_{s}\rightarrow h_1 h_2$ decays in the general
  two-Higgs-doublet model}},
  }{}\href{http://dx.doi.org/10.1103/PhysRevD.64.014014}{Phys.\ Rev.\
  \textbf{D64} (2001) 014014}, \href{http://arxiv.org/abs/hep-ph/0012063}{{\tt
  arXiv:hep-ph/0012063}}\relax
\mciteBstWouldAddEndPuncttrue
\mciteSetBstMidEndSepPunct{\mcitedefaultmidpunct}
{\mcitedefaultendpunct}{\mcitedefaultseppunct}\relax
\EndOfBibitem
\bibitem{Behrens:1998dn}
CLEO collaboration, B.~H. Behrens {\em et~al.},
  \ifthenelse{\boolean{articletitles}}{\emph{{Two-body B meson decays to $\eta$
  and $\eta'$: Observation of $B\to \eta' K$}},
  }{}\href{http://dx.doi.org/10.1103/PhysRevLett.80.3710}{Phys.\ Rev.\ Lett.\
  \textbf{80} (1998) 3710}, \href{http://arxiv.org/abs/hep-ex/9801012}{{\tt
  arXiv:hep-ex/9801012}}\relax
\mciteBstWouldAddEndPuncttrue
\mciteSetBstMidEndSepPunct{\mcitedefaultmidpunct}
{\mcitedefaultendpunct}{\mcitedefaultseppunct}\relax
\EndOfBibitem
\bibitem{PDG2014}
Particle Data Group, K.~A. Olive {\em et~al.},
  \ifthenelse{\boolean{articletitles}}{\emph{{\href{http://pdg.lbl.gov/}{Review
  of particle physics}}},
  }{}\href{http://dx.doi.org/10.1088/1674-1137/38/9/090001}{Chin.\ Phys.\
  \textbf{C38} (2014) 090001}\relax
\mciteBstWouldAddEndPuncttrue
\mciteSetBstMidEndSepPunct{\mcitedefaultmidpunct}
{\mcitedefaultendpunct}{\mcitedefaultseppunct}\relax
\EndOfBibitem
\bibitem{Aubert:2009yx}
BaBar collaboration, B.~Aubert {\em et~al.},
  \ifthenelse{\boolean{articletitles}}{\emph{{B meson decays to charmless meson
  pairs containing $\eta$ or $\eta'$ mesons}},
  }{}\href{http://dx.doi.org/10.1103/PhysRevD.80.112002}{Phys.\ Rev.\
  \textbf{D80} (2009) 112002}, \href{http://arxiv.org/abs/0907.1743}{{\tt
  arXiv:0907.1743}}\relax
\mciteBstWouldAddEndPuncttrue
\mciteSetBstMidEndSepPunct{\mcitedefaultmidpunct}
{\mcitedefaultendpunct}{\mcitedefaultseppunct}\relax
\EndOfBibitem
\bibitem{Schumann:2006bg}
Belle collaboration, J.~Schumann {\em et~al.},
  \ifthenelse{\boolean{articletitles}}{\emph{{Evidence for $B \to \eta'\pi$ and
  improved measurements for $B\to \eta' K$}},
  }{}\href{http://dx.doi.org/10.1103/PhysRevLett.97.061802}{Phys.\ Rev.\ Lett.\
   \textbf{97} (2006) 061802}, \href{http://arxiv.org/abs/hep-ex/0603001}{{\tt
  arXiv:hep-ex/0603001}}\relax
\mciteBstWouldAddEndPuncttrue
\mciteSetBstMidEndSepPunct{\mcitedefaultmidpunct}
{\mcitedefaultendpunct}{\mcitedefaultseppunct}\relax
\EndOfBibitem
\bibitem{Aubert:2008ad}
BaBar collaboration, B.~Aubert {\em et~al.},
  \ifthenelse{\boolean{articletitles}}{\emph{{Measurement of time dependent CP
  asymmetry parameters in $B^0$ meson decays to $\omega K^0_{({\rm S})}$,
  $\eta' K^0$, and $\pi^0 K^0_{({\rm S})}$}},
  }{}\href{http://dx.doi.org/10.1103/PhysRevD.79.052003}{Phys.\ Rev.\
  \textbf{D79} (2009) 052003}, \href{http://arxiv.org/abs/0809.1174}{{\tt
  arXiv:0809.1174}}\relax
\mciteBstWouldAddEndPuncttrue
\mciteSetBstMidEndSepPunct{\mcitedefaultmidpunct}
{\mcitedefaultendpunct}{\mcitedefaultseppunct}\relax
\EndOfBibitem
\bibitem{Chen:2006nk}
Belle collaboration, K.-F. Chen {\em et~al.},
  \ifthenelse{\boolean{articletitles}}{\emph{{Observation of time-dependent CP
  violation in $\decay{\Bz}{\etapr\Kz}$ decays and improved measurements of \CP
  asymmetries in $\Bz\to \phi \Kz, K^0_{\rm S} K^0_{\rm S} K^0_{\rm S}$ and
  $\Bz\to J/\psi\Kz$ decays}},
  }{}\href{http://dx.doi.org/10.1103/PhysRevLett.98.031802}{Phys.\ Rev.\ Lett.\
   \textbf{98} (2007) 031802}, \href{http://arxiv.org/abs/hep-ex/0608039}{{\tt
  arXiv:hep-ex/0608039}}\relax
\mciteBstWouldAddEndPuncttrue
\mciteSetBstMidEndSepPunct{\mcitedefaultmidpunct}
{\mcitedefaultendpunct}{\mcitedefaultseppunct}\relax
\EndOfBibitem
\bibitem{Acciarri:1995bx}
L3, M.~Acciarri {\em et~al.},
  \ifthenelse{\boolean{articletitles}}{\emph{{Search for neutral charmless B
  decays at LEP}},
  }{}\href{http://dx.doi.org/10.1016/0370-2693(95)01042-O}{Phys.\ Lett.\
  \textbf{B363} (1995) 127}\relax
\mciteBstWouldAddEndPuncttrue
\mciteSetBstMidEndSepPunct{\mcitedefaultmidpunct}
{\mcitedefaultendpunct}{\mcitedefaultseppunct}\relax
\EndOfBibitem
\bibitem{Abe:1999ze}
SLAC SLD, K.~Abe {\em et~al.},
  \ifthenelse{\boolean{articletitles}}{\emph{{Search for charmless hadronic
  decays of B mesons with the SLD detector}},
  }{}\href{http://dx.doi.org/10.1103/PhysRevD.62.071101}{Phys.\ Rev.\
  \textbf{D62} (2000) 071101}, \href{http://arxiv.org/abs/hep-ex/9910050}{{\tt
  arXiv:hep-ex/9910050}}\relax
\mciteBstWouldAddEndPuncttrue
\mciteSetBstMidEndSepPunct{\mcitedefaultmidpunct}
{\mcitedefaultendpunct}{\mcitedefaultseppunct}\relax
\EndOfBibitem
\bibitem{Cheng:2009mu}
H.-Y. Cheng and C.-K. Chua, \ifthenelse{\boolean{articletitles}}{\emph{{QCD
  factorization for charmless hadronic $\Bs$ decays revisited}},
  }{}\href{http://dx.doi.org/10.1103/PhysRevD.80.114026}{Phys.\ Rev.\
  \textbf{D80} (2009) 114026}, \href{http://arxiv.org/abs/0910.5237}{{\tt
  arXiv:0910.5237}}\relax
\mciteBstWouldAddEndPuncttrue
\mciteSetBstMidEndSepPunct{\mcitedefaultmidpunct}
{\mcitedefaultendpunct}{\mcitedefaultseppunct}\relax
\EndOfBibitem
\bibitem{Sun:2002rn}
J.~Sun, G.~Zhu, and D.~Du,
  \ifthenelse{\boolean{articletitles}}{\emph{{Phenomenological analysis of
  charmless decays $B_{(s)}\to $ PP, PV, with QCD factorization}},
  }{}\href{http://dx.doi.org/10.1103/PhysRevD.68.054003}{Phys.\ Rev.\
  \textbf{D68} (2003) 054003}, \href{http://arxiv.org/abs/hep-ph/0211154}{{\tt
  arXiv:hep-ph/0211154}}\relax
\mciteBstWouldAddEndPuncttrue
\mciteSetBstMidEndSepPunct{\mcitedefaultmidpunct}
{\mcitedefaultendpunct}{\mcitedefaultseppunct}\relax
\EndOfBibitem
\bibitem{Beneke:2003zv}
M.~Beneke and M.~Neubert, \ifthenelse{\boolean{articletitles}}{\emph{{QCD
  factorization for $ B \to PP$ and $B \to PV$ decays}},
  }{}\href{http://dx.doi.org/10.1016/j.nuclphysb.2003.09.026}{Nucl.\ Phys.\
  \textbf{B675} (2003) 333}, \href{http://arxiv.org/abs/hep-ph/0308039}{{\tt
  arXiv:hep-ph/0308039}}\relax
\mciteBstWouldAddEndPuncttrue
\mciteSetBstMidEndSepPunct{\mcitedefaultmidpunct}
{\mcitedefaultendpunct}{\mcitedefaultseppunct}\relax
\EndOfBibitem
\bibitem{Ali:2007ff}
A.~Ali {\em et~al.}, \ifthenelse{\boolean{articletitles}}{\emph{{Charmless
  non-leptonic $\Bs$ decays to $PP$, $PV$ and $VV$ final states in the pQCD
  approach}}, }{}\href{http://dx.doi.org/10.1103/PhysRevD.76.074018}{Phys.\
  Rev.\  \textbf{D76} (2007) 074018},
  \href{http://arxiv.org/abs/hep-ph/0703162}{{\tt arXiv:hep-ph/0703162}}\relax
\mciteBstWouldAddEndPuncttrue
\mciteSetBstMidEndSepPunct{\mcitedefaultmidpunct}
{\mcitedefaultendpunct}{\mcitedefaultseppunct}\relax
\EndOfBibitem
\bibitem{Williamson:2006hb}
A.~R. Williamson and J.~Zupan, \ifthenelse{\boolean{articletitles}}{\emph{{Two
  body B decays with isosinglet final states in SCET}},
  }{}\href{http://dx.doi.org/10.1103/PhysRevD.74.014003}{Phys.\ Rev.\
  \textbf{D74} (2006) 014003}, \href{http://arxiv.org/abs/hep-ph/0601214}{{\tt
  arXiv:hep-ph/0601214}}\relax
\mciteBstWouldAddEndPuncttrue
\mciteSetBstMidEndSepPunct{\mcitedefaultmidpunct}
{\mcitedefaultendpunct}{\mcitedefaultseppunct}\relax
\EndOfBibitem
\bibitem{Cheng:2014rfa}
H.-Y. Cheng, C.-W. Chiang, and A.-L. Kuo,
  \ifthenelse{\boolean{articletitles}}{\emph{{Updating B\to PP,VP decays in the
  framework of flavor symmetry}},
  }{}\href{http://dx.doi.org/10.1103/PhysRevD.91.014011}{Phys.\ Rev.\
  \textbf{D91} (2015), no.~1 014011},
  \href{http://arxiv.org/abs/1409.5026}{{\tt arXiv:1409.5026}}\relax
\mciteBstWouldAddEndPuncttrue
\mciteSetBstMidEndSepPunct{\mcitedefaultmidpunct}
{\mcitedefaultendpunct}{\mcitedefaultseppunct}\relax
\EndOfBibitem
\bibitem{LHCb-PAPER-2014-026}
LHCb collaboration, R.~Aaij {\em et~al.},
  \ifthenelse{\boolean{articletitles}}{\emph{{Measurement of CP violation in
  $B_s^0 \to \phi \phi$ decays}},
  }{}\href{http://dx.doi.org/10.1103/PhysRevD.90.052011}{Phys.\ Rev.\
  \textbf{D90} (2014) 052011}, \href{http://arxiv.org/abs/1407.2222}{{\tt
  arXiv:1407.2222}}\relax
\mciteBstWouldAddEndPuncttrue
\mciteSetBstMidEndSepPunct{\mcitedefaultmidpunct}
{\mcitedefaultendpunct}{\mcitedefaultseppunct}\relax
\EndOfBibitem
\bibitem{Alves:2008zz}
LHCb collaboration, A.~A. Alves~Jr.\ {\em et~al.},
  \ifthenelse{\boolean{articletitles}}{\emph{{The \lhcb detector at the LHC}},
  }{}\href{http://dx.doi.org/10.1088/1748-0221/3/08/S08005}{JINST \textbf{3}
  (2008) S08005}\relax
\mciteBstWouldAddEndPuncttrue
\mciteSetBstMidEndSepPunct{\mcitedefaultmidpunct}
{\mcitedefaultendpunct}{\mcitedefaultseppunct}\relax
\EndOfBibitem
\bibitem{LHCb-DP-2012-004}
R.~Aaij {\em et~al.}, \ifthenelse{\boolean{articletitles}}{\emph{{The \lhcb
  trigger and its performance in 2011}},
  }{}\href{http://dx.doi.org/10.1088/1748-0221/8/04/P04022}{JINST \textbf{8}
  (2013) P04022}, \href{http://arxiv.org/abs/1211.3055}{{\tt
  arXiv:1211.3055}}\relax
\mciteBstWouldAddEndPuncttrue
\mciteSetBstMidEndSepPunct{\mcitedefaultmidpunct}
{\mcitedefaultendpunct}{\mcitedefaultseppunct}\relax
\EndOfBibitem
\bibitem{Punzi:2003bu}
G.~Punzi, \ifthenelse{\boolean{articletitles}}{\emph{{Sensitivity of searches
  for new signals and its optimization}}, }{} in {\em Statistical Problems in
  Particle Physics, Astrophysics, and Cosmology} (L.~{Lyons}, R.~{Mount}, and
  R.~{Reitmeyer}, eds.), p.~79, 2003.
\newblock \href{http://arxiv.org/abs/physics/0308063}{{\tt
  arXiv:physics/0308063}}\relax
\mciteBstWouldAddEndPuncttrue
\mciteSetBstMidEndSepPunct{\mcitedefaultmidpunct}
{\mcitedefaultendpunct}{\mcitedefaultseppunct}\relax
\EndOfBibitem
\bibitem{LHCb-DP-2014-002}
LHCb collaboration, R.~Aaij {\em et~al.},
  \ifthenelse{\boolean{articletitles}}{\emph{{LHCb detector performance}},
  }{}\href{http://dx.doi.org/10.1142/S0217751X15300227}{Int.\ J.\ Mod.\ Phys.\
  \textbf{A30} (2015) 1530022}, \href{http://arxiv.org/abs/1412.6352}{{\tt
  arXiv:1412.6352}}\relax
\mciteBstWouldAddEndPuncttrue
\mciteSetBstMidEndSepPunct{\mcitedefaultmidpunct}
{\mcitedefaultendpunct}{\mcitedefaultseppunct}\relax
\EndOfBibitem
\bibitem{Skwarnicki:1986xj}
T.~Skwarnicki, {\em {A study of the radiative cascade transitions between the
  Upsilon-prime and Upsilon resonances}}, PhD thesis, Institute of Nuclear
  Physics, Krakow, 1986,
  {\href{http://inspirehep.net/record/230779/files/230779.pdf}{DESY-F31-86-02}}\relax
\mciteBstWouldAddEndPuncttrue
\mciteSetBstMidEndSepPunct{\mcitedefaultmidpunct}
{\mcitedefaultendpunct}{\mcitedefaultseppunct}\relax
\EndOfBibitem
\bibitem{Argus}
ARGUS collaboration, H.~Albrecht {\em et~al.},
  \ifthenelse{\boolean{articletitles}}{\emph{{Search for $b\to s\gamma$ in
  exclusive decays of $B$ mesons}},
  }{}\href{http://dx.doi.org/10.1016/0370-2693(89)91177-5}{Phys.\ Lett.\
  \textbf{B229} (1989) 304}\relax
\mciteBstWouldAddEndPuncttrue
\mciteSetBstMidEndSepPunct{\mcitedefaultmidpunct}
{\mcitedefaultendpunct}{\mcitedefaultseppunct}\relax
\EndOfBibitem
\bibitem{wilks1938}
S.~S. Wilks, \ifthenelse{\boolean{articletitles}}{\emph{The large-sample
  distribution of the likelihood ratio for testing composite hypotheses},
  }{}\href{http://dx.doi.org/10.1214/aoms/1177732360}{Ann.\ Math.\ Statist.\
  \textbf{9} (1938) 60}\relax
\mciteBstWouldAddEndPuncttrue
\mciteSetBstMidEndSepPunct{\mcitedefaultmidpunct}
{\mcitedefaultendpunct}{\mcitedefaultseppunct}\relax
\EndOfBibitem
\bibitem{fsfd}
LHCb collaboration, R.~Aaij {\em et~al.},
  \ifthenelse{\boolean{articletitles}}{\emph{{Measurement of the fragmentation
  fraction ratio $f_s/f_d$ and its dependence on $B$ meson kinematics}},
  }{}\href{http://dx.doi.org/10.1007/JHEP04(2013)001}{JHEP \textbf{04} (2013)
  001}, \href{http://arxiv.org/abs/1301.5286}{{\tt arXiv:1301.5286}}, $f_s/f_d$
  value updated in
  \href{https://cds.cern.ch/record/1559272}{LHCb-CONF-2013-011}\relax
\mciteBstWouldAddEndPuncttrue
\mciteSetBstMidEndSepPunct{\mcitedefaultmidpunct}
{\mcitedefaultendpunct}{\mcitedefaultseppunct}\relax
\EndOfBibitem
\bibitem{LHCb:2012cw}
LHCb, R.~Aaij {\em et~al.},
  \ifthenelse{\boolean{articletitles}}{\emph{{Evidence for the decay $B^0\to
  J/\psi \omega$ and measurement of the relative branching fractions of $B^0_s$
  meson decays to $J/\psi\eta$ and $J/\psi\eta^{'}$}},
  }{}\href{http://dx.doi.org/10.1016/j.nuclphysb.2012.10.021}{Nucl.\ Phys.\
  \textbf{B867} (2013) 547}, \href{http://arxiv.org/abs/1210.2631}{{\tt
  arXiv:1210.2631}}\relax
\mciteBstWouldAddEndPuncttrue
\mciteSetBstMidEndSepPunct{\mcitedefaultmidpunct}
{\mcitedefaultendpunct}{\mcitedefaultseppunct}\relax
\EndOfBibitem
\bibitem{2013lja}
LHCb collaboration, R.~Aaij {\em et~al.},
  \ifthenelse{\boolean{articletitles}}{\emph{{Measurement of the charge
  asymmetry in $B^{\pm}\rightarrow \phi K^{\pm}$ and search for
  $B^{\pm}\rightarrow \phi \pi^{\pm}$ decays}},
  }{}\href{http://dx.doi.org/10.1016/j.physletb.2013.11.036}{Phys.\ Lett.\
  \textbf{B728} (2014) 85}, \href{http://arxiv.org/abs/1309.3742}{{\tt
  arXiv:1309.3742}}\relax
\mciteBstWouldAddEndPuncttrue
\mciteSetBstMidEndSepPunct{\mcitedefaultmidpunct}
{\mcitedefaultendpunct}{\mcitedefaultseppunct}\relax
\EndOfBibitem
\bibitem{Li:2006jv}
H.-n. Li and S.~Mishima,
  \ifthenelse{\boolean{articletitles}}{\emph{{Penguin-dominated $B \to P V$
  decays in NLO perturbative QCD}},
  }{}\href{http://dx.doi.org/10.1103/PhysRevD.74.094020}{Phys.\ Rev.\
  \textbf{D74} (2006) 094020}, \href{http://arxiv.org/abs/hep-ph/0608277}{{\tt
  arXiv:hep-ph/0608277}}\relax
\mciteBstWouldAddEndPuncttrue
\mciteSetBstMidEndSepPunct{\mcitedefaultmidpunct}
{\mcitedefaultendpunct}{\mcitedefaultseppunct}\relax
\EndOfBibitem
\end{mcitethebibliography}

\ifx\mcitethebibliography\mciteundefinedmacro
\PackageError{LHCb.bst}{mciteplus.sty has not been loaded}
{This bibstyle requires the use of the mciteplus package.}\fi
\providecommand{\href}[2]{#2}

\newpage

% Author List ----------------------------                                                                                                                                                                                                                                                                                                
%  You need to get a new author list!                                                                                                                                                                                                                                                                                                    

%%%%%%%%%%%%%%%%%%%%%%%%%%%%%%%%%%%%%%%%%%
\onecolumn
\centerline{\large\bf LHCb collaboration}
\begin{flushleft}
\small

R.~Aaij$^{41}$, 
B.~Adeva$^{37}$, 
M.~Adinolfi$^{46}$, 
A.~Affolder$^{52}$, 
Z.~Ajaltouni$^{5}$, 
S.~Akar$^{6}$, 
J.~Albrecht$^{9}$, 
F.~Alessio$^{38}$, 
M.~Alexander$^{51}$, 
S.~Ali$^{41}$
% ...

{\footnotesize \it
% ...
$ ^{5}$Clermont Universit\'{e}, Universit\'{e} Blaise Pascal, CNRS/IN2P3, LPC, Clermont-Ferrand, France\\
$ ^{6}$CPPM, Aix-Marseille Universit\'{e}, CNRS/IN2P3, Marseille, France\\
% ...
$ ^{9}$Fakult\"{a}t Physik, Technische Universit\"{a}t Dortmund, Dortmund, Germany\\
% ...
$ ^{37}$Universidad de Santiago de Compostela, Santiago de Compostela, Spain\\
$ ^{38}$European Organization for Nuclear Research (CERN), Geneva, Switzerland\\
% ...
$ ^{41}$Nikhef National Institute for Subatomic Physics, Amsterdam, The Netherlands\\
% ...
$ ^{46}$H.H. Wills Physics Laboratory, University of Bristol, Bristol, United Kingdom\\
% ...
$ ^{51}$School of Physics and Astronomy, University of Glasgow, Glasgow, United Kingdom\\
$ ^{52}$Oliver Lodge Laboratory, University of Liverpool, Liverpool, United Kingdom\\
% ...
}
\end{flushleft}
%%%%%%%%%%%%%%%%%%%%%%%%%%%%%%%%%%%%%%%%%%

\newpage

\end{document}